ITEC 625 – Computer Systems Architecture

Research Paper

Hyper Converged Infrastructures: Beyond virtualization.

Student: Alberto Perez Veiga
University of Maryland University College
November 2017

Hyper Converged Infrastructures: Beyond virtualization

## Introduction

Datacenters and their underlaying architecture and infrastructures are in the center of the evolution of technology nowadays. A few years ago, business relayed on error-prone, manual methods to deploy their applications, in many cases completely unaligned with business goals and objectives. These deployments where, in most of the cases, delivered with a stovepipe approach, with a total lack of coherence or awareness on how one could affect each other, without a clear understanding about the relationships among each other. In many cases, technology was used as an end themselves, not as a mean to a purpose, even less as a way to deliver business value. As a consequence, users struggled to effectively align business processes with the technologies provided by the Information Technology (IT) departments, what often caused a complete disconnection between IT and daily operations. Each application was usually installed in a single physical server, without taking into much consideration space, power consumption, maintenance complexity or any other parameter beyond acquisition costs.

With the constant growth of business needs, and the multiplication of applications to fulfil them, the need for additional servers kept growing. Concurrently, IT started to be delivered as a service while hardware kept evolving, becoming more powerful and outgrowing application resource needs. As a consequence, a new trend appeared in the market: virtualization. This was the first phase towards modern datacenters, which helped to consolidate multiple applications in a single server, optimizing resources such as storage, CPU and memory, and provided centralized administration that made IT services management faster and easier.

The rise of new technologies such as e-commerce, Cloud Computing, Big Data or Data Warehousing and Analytics have increased the pressure over the underlying infrastructure to deliver more capabilities using less space, power, cost and complexity. Big companies such as Amazon or Google centralized their massive amount of services in enormous modern datacenters capable of delivering services to a huge number of users. Leveraging on large amounts of bandwidth, processing power and storage capacity, these companies must strive to deliver resilient services to their customers if they want to keep being the top companies in the market.

Many organizations have understood that their future depend on how accurately they align IT with business needs, achieving the most amount of services, of the highest possible reliability and quality, while keeping costs contained. From the first approaches used in datacenters, the IT industry has evolved until reaching the latest evolution in infrastructure: Hyper Converged infrastructures. However, is Hyper Converged for every one? Does it really deliver what it promises or is it only another buzz word?

## Evolution of IT infrastructure

Different literature seems to disagree on the different phases that infrastructure went through since the initial stages. To make things more complicated, different vendors adapt these phases to their particular roadmaps and particular terms. However, there seems to be an agreement in which the path hasn't been exactly linear, and different steps have been taken until the appearance of Hyper Convergence and the concept of Software Defined Datacenter (SDDC). Figure 1 depicts one of the multiple possible taxonomies for this path (West, 2016).

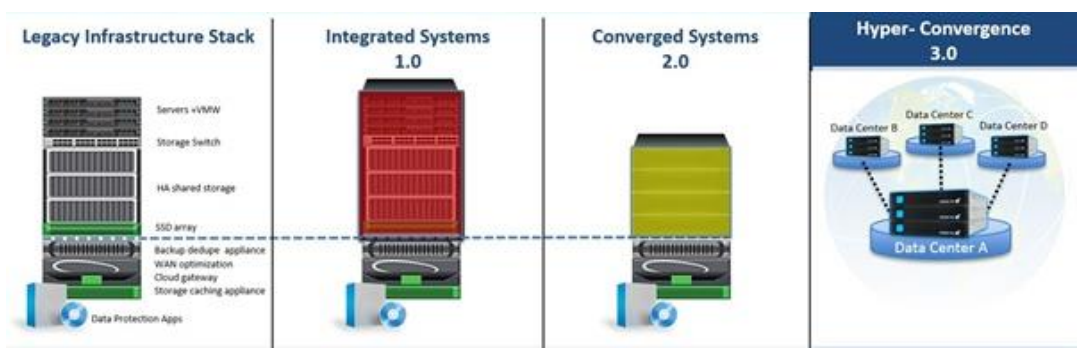

*Figure 1. Path to Hyper Convergence and SDDC (West, 2016)*

Hyper Converged Infrastructures: Beyond virtualization

**Legacy Infrastructure Stack**

The Legacy Infrastructure could be considered a preliminary phase of convergence, where each of the hardware elements of the stack were kept separate and each of them could eventually be provided by a different vendor, leveraging on different possible technologies for their interoperability. Each of the elements were considered as discrete elements, without any kind of software integration among them:

- Servers and Virtualization Hypervisors
- Storage networking
- Shared Storage array (Slow, big disks)
- SSD array (Fast, small Solid-State disks)
- Backup and deduplication array and software
- WAN optimization devices
- Cloud gateways
- Storage optimization appliances
- Data protection / Security Appliances

Although this taxonomy treats as "legacy" this discrete stack of virtualization components, it's worth mentioning that this model is still widely used in multiple datacenters across the world. This was the first unified approach to enterprise virtualization, and it's a trusted model that can leverage on many pieces of existing hardware from different vendors that interoperate based on stablished industry standards and protocols.

The phase named "Integrated Systems" depicts a model where certain components of the stack achieve some level of integration. Although these components still may look like discrete and separated from a physical point of view, several steps were done towards a tighter software integration. This approach tried to achieve an overall facilitated management by bringing together several pieces of software for the management of the different elements.

We could understand this as the first attempt to achieve a consolidated software defined IT management layer. This approach certainly represented an improvement and innovation in terms of deployment times and management easiness. It's worth mentioning that this initial approach had to bring together multiple hardware vendors converging to a single management solution. This approach caused different problems, not only from a technological and interoperability point of view, but also from a political standpoint. Multiple agreements had to be sustained among different companies and cooperation must be close. The fact is that not all companies are willing to give away pieces of their code or collaborate with each other towards a greater good if there is not a clear economic advantage in the deal.

**Convergence**

The third phase of this chronology, named as "Converged systems" depicts a model where part of the aforementioned parts of the stack (Physical Servers, Hypervisor, storage), are consolidated into a single piece from a logical perspective, usually bundled by a certain vendor, who sells it as a suite of hardware and software elements that perform all these functions in a single physical node. Converged systems are usually modular, allowing fast and easy scalability. They allow to stack additional nodes to increase computing power, storage and memory. This approach provides optimized costs, improved speed for deployments and facilitates management. All the nodes are seen and treated by the management software as a unique piece, providing resources as a transparent pool. Networking, storage and computing power are exposed as single management elements rather than the sum of different hardware elements.

Hyper Converged Infrastructures: Beyond virtualizationHyper Converged Infrastructures: Beyond virtualization

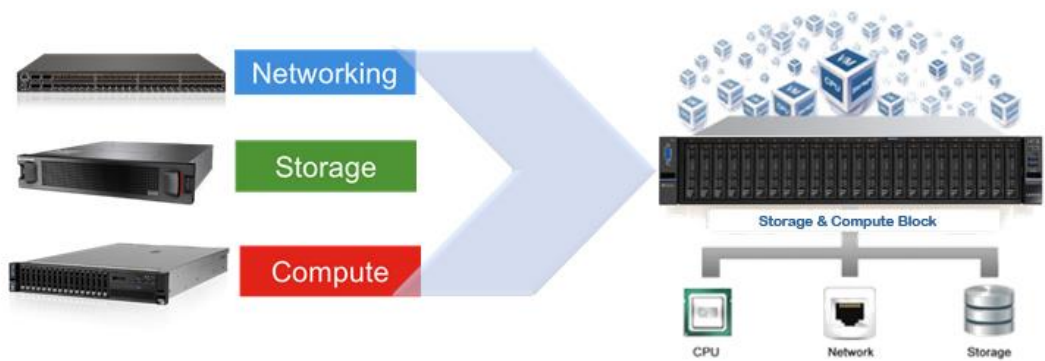

*Figure 2. Converged Infrastructure (Bhattacharya, 2016)*

**Hyper Converged Infrastructure (HCI)**

The final phase, one of the trending topics in the IT industry right now, is Hyper Convergence. In this model, infrastructure is strongly software-centric, with closely integrated computing, networking, virtualization and storage resources, as opposed to the traditional (legacy) infrastructure, where most of these resources were separate physical elements. This kind of infrastructure doesn't only focus on centralizing and compacting the local datacenter, but also to integrate and consolidate management and resources from remote datacenters, extending the management and administration beyond the boundaries of the local facilities.

Hyper Converged Infrastructures allow all their integrated technologies to be managed as a single system using a single, centralized management software, while allowing the addition of new capacities expanding the base systems with additional nodes. In general, HCI can be defined as a virtualized computing infrastructure that combines the services of the data center in an appliance form factor. This approach is supposed to accelerate the speed of virtualized deployments, reduce their complexity, improve operations and lowers costs.

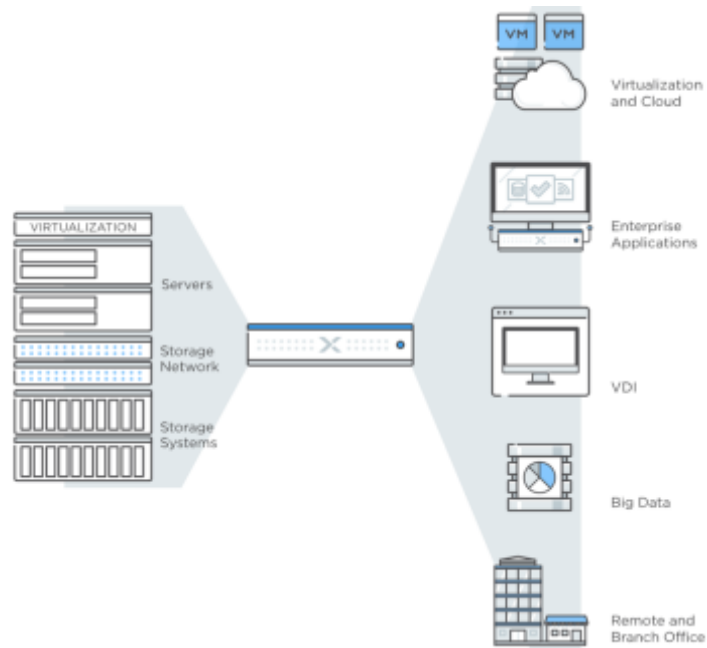

*Figure 3. Hyper Converged Node (Nutanix)*

HCI provides, over all, a software-centric design model, where all the hardware layers are abstracted in favor of a higher-level view from a management perspective. The whole architecture is conceived using appliances as building-blocks which, when combined, scale to achieve the necessary capacity. This philosophy, which encompasses a single vendor as provider, offers high levels of automation and efficient aggregation of additional components. The fact that all of this building blocks are provided by the same vendor, also improves support levels and eliminates compatibility and interoperability issues among components. HCI eliminates infrastructure silos, avoiding the hidden costs of systems that are purchased for a certain purpose and cannot be further integrated in the overall schema of the data center.

At this point and within this context, it's also interesting to mention the concept of Software Defined Datacenter (SDDC) and the relationship among SDDC and HCI. SDDC can be defined as a data facility where all his components have been virtualized: computing, storage, networking, security, management, etc. SDDC, could be considered an evolution of HCI in which the latter would be only a part of it.

In general, the adoption of HCI or ultimately its evolution to SDDC provides a virtualization-ready environment, highly efficient and scalable, which results from a paradigm shift from a hardware-centric to an application-centric approach. The ultimate benefit of HCI is not only a reduced Capital Expenditure (Capex) in the acquisition of the initial capability, but also a reduced Operations Expenditure (Opex) as a result of the simplified scalability model.

**Costs**

Studies point out that around 70% of the typical IT budget is spent on maintenance costs of enterprise data centers (Lawton, 2014). The more efficient installation of patches and upgrades with HCI is supposed to reduce maintenance and administration costs. One of the problems of legacy systems is the adoption of different parts in silos, which causes incremental costs increases. The transition to HCI should provide costs savings over time, due to the mentioned reduced times of exploitation, reduced needs for maintenance, deployment, personnel training, etc. Although for a number of companies it may be needed to retain old hardware for particular legacy applications, the migration to this new approach is supposed to be cost-beneficial (Lawton, 2014).

The big cost savings seem to come, though, from the operational expenses, due mostly to the commodity hardware used for the base implementation and further expansion of the nodes, which streamlines the scaling of the infrastructure when new needs arise over time and the infrastructure has to be expanded (Lawton, 2014).

Hyper Converged Infrastructures: Beyond virtualization

## The reality of HCI

The previous sections of this research have described the theory behind HCI, its concepts and the evolution of the IT infrastructure industry approach from the era of a physical server per application until SDDC (paradigm shift from hardware-centric to application-centric). However, it has to be taken into account that the world of IT is often driven by marketing, high-level approaches without clear real implementations and architectural visions without feasible implementation. During the last few years, almost every company seems to have came up with their own HCI product. The term "Hyper Convergence" is risking to become the next buzz word of the IT industry, as some companies seem to be approaching HCI as a meaningless bundle of commodity hardware, with a common re-branded, in-house tailored hypervisor and some sort of software-defined networking approach (Toigo, 2015). Although this approach doesn't seem to be in line with the theoretical concept and main tenets of this technology, these solutions are labelled as "Convergent" due to the simple fact that they are bundled together and sold by the same vendor.

As the whole concept behind HCI and SDDC relies in a software-defined architecture, seems advisable to emphasize the need to include robust and mature software in the implementation, that should be used to manage high-quality hardware, properly tested and bundled together to provide reliable products. Some initial approaches involving Software-Defined Storage (SDS) technologies such as VMware VSan, which integrates SDS in a Hypervisor-centric approach, looks promising. Additionally, there are interesting evolutions towards the SDDC concept such as VMware EVO: RAIL, which includes the well-known VMware software stack, bundled on the top of top-of-the-line hardware vendors.

As the costs of adopting one of these vendor's proposals can be initially high, other companies take a different approach and, surfing the wave of the promises of reduced costs, try to cut corners delivering products that seem to be re-brands of their own hardware, bundled with different also re-branded pieces of software (often open-source), which they claim to be their own product. Nothing illegal behind this approach, besides the doubtful promise of performance and quality to customers looking for bargains, which may very well end-up tied-up to a vendor which doesn't deliver the expected quality, support, or frequency of updates and security patches required in today's rapid evolving IT scene.

  One of the advantages of HCI is the solid compatibility of the components integrating the solution. This benefit comes, however, with a trade-off: performance. The cost of compatibility is a massive need for interoperability testing among all the components of the solution that usually takes months. This amount of testing causes that, when the solution reaches the market, some of the components may not be the most recent. So, if the infrastructure needs the greatest performance and the latest hardware technologies, HCI is probably not a good approach to start with (Harvey, 2017).

  Although HCI promises costs reduction in the long run, it's important to carefully plan ahead and understand future needs within the roadmap of the organizational business strategy. The initial investment may be expensive and, being these systems based on commodity hardware, some vendors charge also expensive premiums for support contracts. If the customer is on his own, without a support contract, troubleshooting activities can be complicated due to the tight integration of all the components of the solution, which will make difficult to understand if problems are coming from network, storage, computing or software components.

  Companies providing HCI promise flexible, cost-effective scaling. However, when adopting HCI, organizations should take into account the medium and long-term scaling needs. As an example, if the systems host a data-intensive application, the near future need may be to increment storage capacity. Given the fact that HCI doesn't allow to expand only the storage component, the upgrade will probably end up in an expensive scaling with the need to add additional HCI nodes whereas the legacy model would have allowed to add only storage, resulting in a cheaper upgrade.

Last, but not least, the essence of HCI, a software-centric bundle, sold and supported by a single vendor, is a disadvantage in itself. Once one of these platforms is implemented, it will most likely be difficult to integrate or move to a different vendor in the infrastructure without replacing the existing hardware or going through a major refurbishing. As all the hardware is intended to run under a common management software, it will be complicated to make it run together with a product from a different vendor or under a different software.

## Conclusion

HCI has brought virtualization and IT strategies to a new level. Datacenters are undergoing a deep paradigm shift from a hardware-centric to an application-centric approach which leverages on software defined architectures, while IT is more and more being delivered as services rather than assets or products. Throughout different evolving phases since the initial attempts to convergence, the concept has been refined down to a level where, ultimately, a whole datacenter could be fully managed from a centralized single point, abstracting the whole hardware layer and exposing it to the administrators as a transparent pool of resources.

HCI promises to deliver cheaper services in less time, using less physical space, with lower power consumption and minimal management needs, reducing overall costs. As a hot topic in the IT industry, many companies are jumping into the opportunity, delivering their own implementations of the concept, surfing the wave of the new trend. However, there are several back draws and trade-offs in terms of performance, hidden costs, future compatibility, support, scalability and security, among others.

   Organizations should avoid, in general, moving blindly to HCI. As any other IT strategy, the migration to an HCI platform should take into account the medium and long term business goals and objectives and how these may require IT to adapt to them. Initial acquisition costs of these solutions can be expensive and, depending on the kind of applications used, scalability may be expensive if only certain parts of the stack need to be upgraded.

   IT industry is strongly marketing driven. Many vendors understand it clearly and leverage on the traction of buzz words and the hype created behind them, delivering products of doubtful quality. Although HCI seems to start being a stablished trend, rather than just a hype, the smart customer should strive to understand the theoretical basis and concepts behind it, plan ahead always within the scope of business strategies, and only implement solutions provided by trusted vendors that really fulfil their needs.

Hyper Converged Infrastructures: Beyond virtualization